\documentclass[twocolumn,showpacs,prl]{revtex4}
\usepackage{graphicx}

\newcommand{\ket}[1]{| #1 \rangle}
\newcommand{\bra}[1]{\langle #1 |}
\newcommand{\braket}[2]{\langle #1 | #2 \rangle}

\begin{document}
\title{Unconditional security of coherent-state quantum key distribution with strong phase-reference pulse}

\author{Masato Koashi}
\affiliation{CREST Research Team for Photonic Quantum Information, School
of Advanced Sciences,\\ The Graduate University for  Advanced Studies
(SOKENDAI), Hayama, Kanagawa, 240-0193, Japan}

\begin{abstract} 
We prove the unconditional security of a quantum key distribution 
protocol in which bit values are encoded in the phase of a weak 
coherent-state pulse relative to a strong reference pulse. 
In contrast to implementations in which a weak pulse
is used as a substitute for a single-photon source,  
the achievable key rate is found to decrease only linearly 
with the transmission of the channel. 
\pacs{03.67.Dd 03.67.-a}
\end{abstract}

\maketitle

Quantum key distribution provides a way to distribute 
a secret key between two distant parties, Alice and Bob,
even if the quantum channel between them suffers from 
small noises. As long as the law of quantum
mechanics is valid, an eavesdropper, Eve, cannot force Alice 
and Bob to accept a key on which she has a 
nonnegligible amount of information.
A proof of such unconditional security was first provided
by Mayers \cite{Mayers96} for the BB84 protocol \cite{Bennett-Brassard84},
followed by other proofs
\cite{others,Shor-Preskill00,ILM01,GLLP02,TKI03,Tamaki-Lutkenhaus04}.  
While a perfect single-photon source is assumed in the earlier
proofs, recent proofs \cite{ILM01,GLLP02} cover the use of 
a weak laser pulse in a coherent state as a substitute for 
a single photon. This is good news in the practical point of view,
but comes with a price: the multiphoton components of the weak 
pulse allow Eve a so-called photon-number splitting 
attack \cite{Lutkenhaus00,BLMS00}. 
In order to achieve the security under this attack,
Alice must lower the amplitude of her weak pulse as the loss in  
the channel increases. As a result, there is a bound 
\cite{Lutkenhaus00} on 
the achievable key rate which scales as $O(\eta^2)$ with 
channel transmission $\eta$.

In this paper, we prove an unconditional security of a
scheme using a weak coherent pulse and achieving a rate 
that scales as $O(\eta)$. The scheme is essentially the one
proposed by Bennett \cite{Bennett92}, in which a strong
pulse is transmitted as a phase reference together with a 
weak pulse containing the bit information in the relative 
phase. We made a minor modification to introduce a second 
local oscillator (LO) for Bob. This makes the analysis simpler, 
and allows us to assume a realistic threshold detector that may 
be noisy, inefficient, sensitive to multimodes of light, and
only discriminates the vacuum from one or more photons.

\begin{figure}[tbp]
\begin{center}
 \includegraphics[scale=1.000]{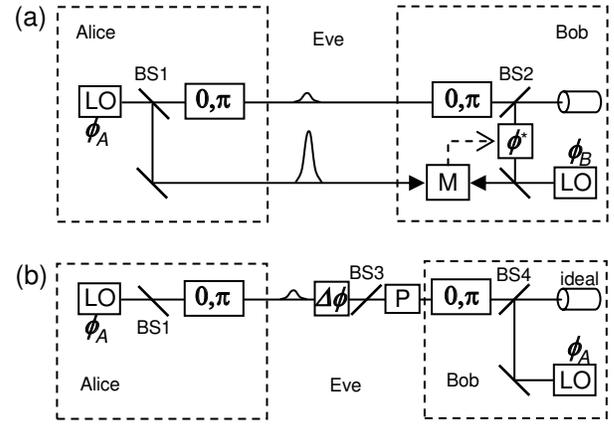}
 \end{center}
 \caption{(a) A scheme with a strong reference pulse.
(b) An equivalent scheme except that Eve's region is extended.
 \label{schematic}}
\end{figure}

The scheme is depicted in Fig.~\ref{schematic}(a). 
Suppose that Alice's LO emits a strong pulse in 
a coherent state
 with complex amplitude $|\alpha_0|e^{i\phi_A}$.
Using an asymmetric beamsplitter (BS1), Alice extracts a 
weak pulse 
with very small amplitude $\alpha=|\alpha|e^{i\phi_A}$, and 
encodes a randomly chosen bit value 
$0$ or $1$
by applying phase shift $0$ or $\pi$, resulting in state $\ket\alpha$ or
$\ket{-\alpha}$, respectively. 
Together with this signal, she sends the strong pulse from the other output of
BS1 to Bob as a phase reference.

On the receiver's side, Bob chooses randomly a bit value
$0$ or $1$, and applies phase shift $0$ or $\pi$ to the 
weak signal pulse, respectively. Instead of using the reference 
pulse from Alice directly, Bob uses another LO and tries to lock 
its phase to Alice's one.
Suppose that Bob's LO produces a strong
pulse with complex amplitude 
$|\beta_0|e^{i\phi_B}$. Combining a potion of this pulse and 
the reference pulse from Alice, he conducts a series of interference 
experiments (M) to infer the phase difference $\phi_A-\phi_B$. He then 
applies a phase shift equal to 
this estimated value $\phi^*$ to his LO, and mixes it with the weak signal 
from Alice at BS2. The mixed signal is measured by a threshold
detector, which gives a ``click'' whenever it receives one or more
photons. Bob reports the outcome of the detector to Alice over an
authenticated public channel. The click implies a conclusive result,
 and both parties accept their bits. No click implies an inconclusive 
result, and they discard the bits.

The security analysis in this paper is valid even if LOs with
phases $\phi_A$ and $\phi_B$ are available to Eve. 
Then, the reference pulse from Alice 
gives no information to Eve. The only effect of Eve's attack on this pulse 
is to disturb the measurement outcome $\phi^*$ to be deviated from the 
desired value, as $\phi^*=\phi_A-\phi_B-\Delta\phi$. But exactly the
same effect can be obtained by just applying the phase shift $\Delta\phi$
to the weak signal from Alice (Eve may simulate M by herself). 
Hence we can safely assume that Eve
simply ignores the strong reference pulse. Similarly, any imperfection 
in the estimation process M, including the fundamental limitation
arising from finiteness of the amplitudes of the two LOs, has the same
effect as introducing a noise source applying a phase shift $\Delta\phi$
on the weak signal while assuming a perfect estimation, 
$\phi^*=\phi_A-\phi_B$.

The major imperfections in the detector can be treated as follows.
Suppose that the quantum efficiency of the detector is $\eta_{\rm D}$,
the transmission coefficient of BS2 is $\eta_{\rm BS2}$, and 
the amplitude of LO incident on BS2 is 
$(1-\eta_{\rm BS2})^{-1}\eta^{-1}_{\rm D}\beta$. Then, the same
measurement can be implemented by inserting a lossy medium 
(BS3) with transmission
$\eta_{\rm BS2}\eta_{\rm D}\eta^{-1}_0$, then mixing LO with amplitude 
$(1-\eta_0)^{-1}\beta$ by a beamsplitter BS4 
with transmission $\eta_0$, followed by a
detector with unit efficiency. Here we take the limit of
$\eta_0\rightarrow 1$. The dark counting of the detector or 
the detection of stray photons can be simulated by a device (P) that 
inserts a photon in a mode that is orthogonal to the modes of the LOs.
We thus finally arrive at a scheme with an ideal threshold detector and 
a locked pair of LOs, as in Fig.~\ref{schematic}(b). In this
figure, the region accessible by Eve is extended for the sake of
simplicity. If a protocol is secure with this scheme, the same
protocol implemented by the scheme in Fig.~\ref{schematic}(a) is also secure.

Bob's decision process in the scheme in Fig.~\ref{schematic}(b)
can be regarded as a 
generalized measurement on the light entering his site
with three outcomes, $0$, $1$, and $2$, where the last one means
``inconclusive''. Let 
${\cal H}_B={\cal H}_0\otimes {\cal H}_1\otimes 
\cdots \otimes {\cal H}_\nu \otimes\cdots$ be the Hilbert space for the light
modes received by Bob that are sensible by the detector.
The mode $\nu=0$ represents the pulse mode of Bob's LO,
and the modes with $\nu\ge 1$ are orthogonal to it.
Let us write the coherent state
$\ket{\beta}_0\ket{0}_1\ket{0}_2\cdots$ simply as $\ket{\beta}$.
Then, the generalized measurement is described by the POVM
$\{F_0,F_1,F_2\}$, where 
$$
F_0=({\bf 1}-\ket{-\beta}\bra{-\beta})/2, \;\;
F_1=({\bf 1}-\ket{\beta}\bra{\beta})/2,
$$
and $F_2={\bf 1}-F_0-F_1$. If everything is ideal except for the transmission 
$\eta$ in the channel, Alice's signal is received by Bob in coherent states 
$\ket{\pm \sqrt{\eta}\alpha}$, and they can agree on a key without errors 
by choosing $\beta=\sqrt{\eta}\alpha$.

Before describing the proof of unconditional security, we 
introduce several notations. We decompose 
${\cal H}_B$ as ${\cal H}_B={\cal K}_B\oplus {\cal H}_{\rm ex}$,
where ${\cal K}_B$ is the two-dimensional subspace spanned by
$\ket{\beta}$ and $\ket{-\beta}$. We assume 
$\alpha$ and $\beta$ to be real and positive without loss of generality.
Let $\{\ket{\mu_l}_B\}_{l=1,2,\ldots}$
be an arbitrary complete orthonormal basis for ${\cal H}_{\rm ex}$.
We identify ${\cal K}_B$ as a qubit, and define its $X$ basis
as $\{\ket{0_x}_B\equiv (\ket{\beta}+\ket{-\beta})/(2c_\beta),
\ket{1_x}_B\equiv (\ket{\beta}-\ket{-\beta})/(2s_\beta)\}$,
where $2c_\beta^2-1\equiv 1-2s_\beta^2\equiv\braket{-\beta}{\beta}
=e^{-2|\beta|^2}$. The $Z$-basis states are denoted as 
$\ket{j_z}_B\equiv (\ket{0_x}_B+(-1)^j\ket{1_x}_B)/\sqrt{2}$ $(j=0,1)$.
For Alice's side, we denote by ${\cal H}_A$ the Hilbert space of the
light modes emitted from her site. 
We also introduce an auxiliary qubit in Alice's site, with Hilbert 
space ${\cal K}_A$. We denote the $X$- and the $Z$-basis states as 
$\ket{j_x}_A$ and $\ket{j_z}_A$ $(j=0,1)$. We sometimes denote
the projection $\ket{\Phi}\bra{\Phi}$ as $P(\ket{\Phi})$.

The key idea in the security proof is a trace-nonincreasing 
completely positive map, which is specified by Kraus operators 
$A_j:{\cal H}_B\rightarrow {\cal K}_B$ $(j=0,1,2,\ldots)$ defined by
$A_0=s_\beta\ket{0_x}_B\bra{0_x}+c_\beta\ket{1_x}_B\bra{1_x}$ for
$j=0$
and $A_j=\ket{0_x}_B\bra{\mu_j}$ otherwise.
Since $\sum_j A_j^\dagger A_j\le 1$, there exists
a filter with the following property. It takes any state $\rho$ acting on 
${\cal H}_B$ as an input, and it accepts with probability 
$p=\sum_j {\rm Tr}(A_j^\dagger A_j\rho)$ while it rejects 
with probability $1-p$. 
Whenever it accepts, it returns the output state 
$\sum_j A_j\rho A_j^\dagger/p$ acting on ${\cal K}_B$.
This filter is related to the POVM $\{F_0,F_1,F_2\}$ by
\begin{equation}
F_k=\sum_j A_j^\dagger \ket{k_z}_B\bra{k_z} A_j
\label{filter-POVM}
\end{equation}
for $k=0,1$, which is easily confirmed.
This relation implies that we can implement 
the measurement $\{F_0,F_1,F_2\}$ by applying the filter and conducting 
$Z$-basis measurement on the output state when it accepts (if it
rejects, we assume that the outcome is ``2''). 

With the above decomposition of Bob's measurement, we
can prove the unconditional security by a method similar 
to the cases of qubit-based B92 protocols \cite{TKI03,Tamaki-Lutkenhaus04}.
We introduce a protocol based on entanglement distillation \cite{BDSW96}, 
which is later shown to be equivalent to the real protocol.
In the new protocol, (1) Alice prepares 
state 
$(\ket{0_z}_A\ket{\alpha}+\ket{1_z}_A\ket{-\alpha})/\sqrt{2}$ on
${\cal K}_A\otimes {\cal H}_A$.
We assume that Alice produces $2N$ copies of this state.
(2) Eve receives $2N$ pulses (corresponding to ${\cal H}_A^{\otimes 2N}$) 
from Alice, and prepares a state on ${\cal H}_B^{\otimes 2N}$, which 
may be entangled to Eve's system. (3) After Bob has received 
$2N$ pulses (corresponding to ${\cal H}_B^{\otimes 2N}$), 
Alice and Bob randomly permutate the order of $2N$ pairs of systems 
by public discussion. (4) For the first $N$ pairs (check pairs), 
Alice measures each qubit (${\cal K}_A$) on $Z$ basis, 
and Bob performs the POVM $\{F_0,F_1,F_2\}$
on each pulse (${\cal H}_B$). They disclose
all the results, and learn the number $n_{\rm err}$ of 
error events where the combination of 
Alice's and Bob's outcomes are $(0,1)$ or $(1,0)$.
(5) For the other $N$ pairs (data pairs),
Bob applies the above filter to each pulse,
and discloses each result (accept or reject).
Let $n_{\rm fil}$ be the number of events where 
the filter has accepted. (6) Alice and Bob now 
have $n_{\rm fil}$ pairs of qubits 
(${\cal K}_A \otimes {\cal K}_B$), from which 
they try to extract a number of pairs in the 
maximally entangled state 
$(\ket{0_z}_A\ket{0_z}_B+\ket{1_z}_A\ket{1_z}_B)/\sqrt{2}$.
To do so, they estimate the number $n_{\rm bit}$ of pairs 
with a bit error (represented by the subspace spanned by
$\{\ket{0_z}_A\ket{1_z}_B,\ket{1_z}_A\ket{0_z}_B\}$)
and the number $n_{\rm ph}$ of pairs with a phase error
(the subspace spanned by 
$\{\ket{0_x}_A\ket{1_x}_B,\ket{1_x}_A\ket{0_x}_B\}$),
from the knowledge of $n_{\rm fil}$ and $n_{\rm err}$.
If neither number of errors is too high, they run an 
entanglement distillation protocol (EDP) and then measure 
on $Z$ basis to determine the final key.
As in the proof of BB84 \cite{Shor-Preskill00}, if the estimation of
the upper bounds for $n_{\rm bit}$ and $n_{\rm ph}$
is correct except for a probability that becomes 
exponentially small as $N$ increases, this protocol
is essentially secure.

According to the argument by Shor and Preskill \cite{Shor-Preskill00}, 
if we choose an appropriate EDP scheme,
Alice and Bob can conduct $Z$-basis measurement
on the $n_{\rm fil}$ pairs immediately after 
step (5) and decide the final key by 
a public discussion without compromising 
 the security.
Then, Eq.~(\ref{filter-POVM}) shows that 
Bob's measurement on each data qubit is also 
the POVM $\{F_0,F_1,F_2\}$. 
Alice's measurement can be further brought 
forward to the end of step (1), then this step
is equivalent to just preparing state 
$\ket{\alpha}$ or $\ket{-\alpha}$ randomly.
The new protocol is thus equivalent to the 
prepare-measure protocol implemented as in 
Fig.~\ref{schematic}(b).

The remaining task for the security proof is to establish 
an exponentially good way of estimating $n_{\rm bit}$ and 
$n_{\rm ph}$. Since $n_{\rm err}$ and $n_{\rm bit}$ are
 the results of the same measurement applied to 
the (randomly assigned) check pairs and 
to the data pairs, we can apply a classical probability
estimate to see that
$|n_{\rm bit}-n_{\rm err}|\le N\epsilon$ 
holds except for a small probability which 
is asymptotically smaller than $\sim \exp(-N\epsilon^2)$.
The estimation of $n_{\rm ph}$ can be done by considering 
what could have happened if Alice and Bob measured their 
$n_{\rm fil}$ pairs of data qubits in $X$ basis and determined 
$n_{\rm ph}$ by discussion, just after the step (5). 
In this scenario, they obtain three numbers 
$(n_{\rm fil},n_{\rm ph},n_{\rm err})$. 
The following argument shows that some combinations 
of $(n_{\rm fil},n_{\rm ph},n_{\rm err})$ are exponentially
rare for any attack by Eve, and hence gives an 
(exponentially reliable) upper bound 
$\bar{n}_{\rm ph}(n_{\rm fil},n_{\rm err})$
for $n_{\rm ph}$ as a function of the other two.

We can regard $n_{\rm ph}$ as the number of events 
where a measurement on ${\cal K}_A\otimes {\cal H}_B$
produced the outcome corresponding to the element of a POVM
$M_{\rm ph}\equiv \sum_j P(\ket{0_x}_A)\otimes A_j^\dagger
\ket{1_x}_B\bra{1_x}A_j+P(\ket{1_x}_A)\otimes A_j^\dagger
\ket{0_x}_B\bra{0_x}A_j=s_\beta^2P(\ket{1_x}_A\ket{0_x}_B)
+c_\beta^2P(\ket{0_x}_A\ket{1_x}_B)+
P(\ket{1_x}_A)\otimes {\bf 1}_{\rm ex}$.
Similarly, $n_{\rm fil}$ corresponds to 
$M_{\rm fil}\equiv {\bf 1}_A\otimes  \sum_j A_j^\dagger A_j
={\bf 1}_A\otimes(c_\beta^2P(\ket{1_x}_B)
+s_\beta^2P(\ket{0_x}_B)+{\bf 1}_{\rm ex})$.
From these forms, we notice that $n_{\rm ph}$ and $n_{\rm fil}$
are also obtained by the projection measurement
$\{P_{00}, P_{11},P_{10}, P_{01},
P(\ket{0_x}_A)\otimes {\bf 1}_{\rm ex},P(\ket{1_x}_A)\otimes 
{\bf 1}_{\rm ex}\}$,
where $P_{ij}\equiv P(\ket{i_x}_A\ket{j_x}_B)$, followed 
by a classical procedure composed of Bernoulli trials.
If we denote the results of the $N$ projection measurements
as $\{n_+(1-\delta_+),n_+\delta_+,n_-(1-\delta_-),n_-\delta_-,
m_0,m_1\}$
in the same order,
these numbers should be related to $n_{\rm ph}$ and $n_{\rm fil}$ as
\begin{eqnarray}
|n_{\rm ph}-m_1-n_-[s_\beta^2(1-\delta_-)+c_\beta^2\delta_-]|&\le& 
N\epsilon
\\
|n_{\rm fil}-m_0-m_1-c_\beta^2(n_+\delta_++n_-\delta_-)&&
\nonumber \\
-s_\beta^2 [(n_+(1-\delta_+)+n_-(1-\delta_-)]|&\le &
N\epsilon
\end{eqnarray}
with probability at least $1-\exp(2N\epsilon^2)$. Since the marginal 
state $\rho_A$ on ${\cal K}_A$ cannot be altered by Eve, 
the $X$-basis measurement on ${\cal K}_A$ is another Bernoulli trial.
Since $\rho_A=c_\alpha^2P(\ket{0_x}_A)+s_\alpha^2P(\ket{1_x}_A)$,
we have 
$$
|m_1+n_+\delta_++n_-(1-\delta_-)-s_\alpha^2N|\le N\epsilon.
$$

For the check pairs, $n_{\rm err}$ corresponds to 
$M_{\rm err}\equiv P(\ket{0_z}_A)\otimes F_1+P(\ket{1_z}_A)\otimes F_0
=(1/2)[P(\ket{\Gamma_{11}})+P(\ket{\Gamma_{01}})
+{\bf 1}_A\otimes {\bf 1}_{\rm ex}]$, where we have introduced a basis 
$\{\ket{\Gamma_{ij}}\}_{i,j=0,1}$ of ${\cal K}_A\otimes {\cal K}_B$
by $\ket{\Gamma_{ij}}\equiv c_\beta\ket{i_x}_A\ket{j_x}_B
-(-1)^j s_\beta\ket{(1-i)_x}_A\ket{(1-j)_x}_B$.
It implies that $n_{\rm err}$ could also be obtained by
the global projection measurement
$\{Q_{00}, Q_{11},Q_{10}, Q_{01},
{\bf 1}_A\otimes {\bf 1}_{\rm ex}\}$, where $Q_{ij}\equiv P(\ket{\Gamma_{ij}})$,
 followed by Bernoulli trials.
If we write the results of $N$ projection measurements
as $\{n'_+(1-\delta'_+),n'_+\delta'_+,n'_-(1-\delta'_-),n'_-\delta'_-,
m\}$, we obtain
$$
|n_{\rm err}-(n'_+\delta'_++n'_-\delta'_-+m)/2|\le N\epsilon.
$$

\begin{figure}[tbp]
\begin{center}
 \includegraphics[scale=0.450]{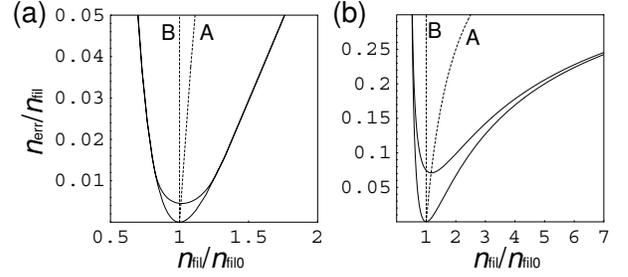}
 \end{center}
 \caption{Key gain $G$ is positive in the region between the two solid
 curves. (a) $|\alpha|^2=0.5$, $\eta=0.01$, (b)
$|\alpha|^2=10^{-3}$, $\eta=0.01$.
\label{region}}
\end{figure}

 If we compare the projection measurements on the data pairs and the check
 pairs, we 
further notice that $n_+$ and $n'_+$ are the results of an identical
measurement, namely, projection onto the space ${\cal H}_+$ spanned by
$\{\ket{0_x}_A\ket{0_x}_B,\ket{1_x}_A\ket{1_x}_B\}$. We can thus apply
 the classical probability estimate.
$\delta_+$ and $\delta'_+$ comes from projection to nonorthogonal
 states. For such a case, it was shown \cite{TKI03} that 
combination $(\delta_+,\delta'_+)$ is exponentially rare 
unless there exists a state $\rho$ on ${\cal H}_+$ satisfying
${\rm Tr}[\rho P(\ket{1_x}_A\ket{1_x}_B)]=\delta_+$
and ${\rm Tr}[\rho P(\ket{\Gamma_{11}})]=\delta'_+$.
Using these arguments, we obtain
\begin{eqnarray}
&&|n_\pm-n'_\pm|\le N\epsilon,
\nonumber \\
\delta'_\pm\ge c_\beta^2\delta_\pm&+&s_\beta^2(1-\delta_\pm)
-2c_\beta s_\beta\sqrt{\delta_\pm(1-\delta_\pm)}-\epsilon'.
\nonumber
\end{eqnarray}

We are interested in the secret key gain in the limit 
$N\rightarrow \infty$.
Setting $\epsilon$ and $\epsilon'$ to be zero, we obtain 
$2n_{\rm err}\ge n_{\rm fil}-
2c_\beta s_\beta[n_+\sqrt{\delta_+(1-\delta_+)}+
n_-\sqrt{\delta_-(1-\delta_-)]}$. Then we can eliminate
$n_\pm$ and $\delta_\pm$ to be left with two free parameters
$m_0$ and $m_1$. From this point, in general, we may 
have to numerically minimize $n_{\rm err}$ over 
the two parameters. It turned out that in most of 
 interesting cases $m_0=m_1=0$ gives the minimum.
Once we obtain the minimum of $n_{\rm err}$ as a 
function of $(n_{\rm ph}, n_{\rm fil})$, 
we can determine $\bar{n}_{\rm ph}(n_{\rm fil},n_{\rm err})$.
The length of the final key is given \cite{Shor-Preskill00,GLLP02} by 
$n_{\rm key}(n_{\rm fil},n_{\rm err})=
n_{\rm fil}[1-h(n_{\rm err}/n_{\rm fil})-h(\bar{n}_{\rm ph}/n_{\rm
fil})]$ when this value is nonnegative and 
$2 \bar{n}_{\rm ph}\le n_{\rm fil}$.

Figure \ref{region} shows the parameter region 
$(n_{\rm fil},n_{\rm err})$ where the key gain 
$G\equiv n_{\rm key}/N$ is positive,
for a few choices of $\alpha$ and $\beta=\sqrt\eta\alpha$.
 When Alice chooses 
$|\alpha|^2=0.5$, the tolerable error rate 
$n_{\rm err}/n_{\rm fil}$ is less than 1\%.
For a smaller amplitude $|\alpha|^2=0.001$, the tolerable
rate increases to $\sim 7\%$. Choosing a smaller value for 
$|\alpha|^2$ than 
this example does not improve the tolerable rate significantly.
For either case in Fig.~\ref{region}, taking a smaller value 
of $\eta$ gives little change in the shape of region, except for 
the normalization factor in the abscissa 
$n_{\rm fil0}\equiv N(1-e^{4\eta|\alpha|^2})/2\sim 2\eta|\alpha|^2N$.
This allows us to choose a fixed $|\alpha|^2$ in the limit of 
$\eta\rightarrow 0$ as long as $n_{\rm err}/n_{\rm fil}$ is fixed, 
leading to the key gain $G$ proportional to $\eta$.

In Fig.~\ref{region}(b), we notice that the region 
extends far into the area with high 
$n_{\rm err}/n_{\rm fil}$ and $n_{\rm fil}$, but ordinary 
sources of errors never achieve this region. For example,
Errors in phase [$\Delta\phi$ in Fig.~\ref{schematic}(b)]
result in curve B. Errors by spurious countings 
[device P in Fig.~\ref{schematic}(b)], which is modeled as
$n_{\rm fil}=N\lambda+(1-\lambda)n_{\rm fil0}$ 
and $n_{\rm err}=N\lambda/2$, follow curve A.

\begin{figure}[tbp]
\begin{center}
 \includegraphics[scale=0.800]{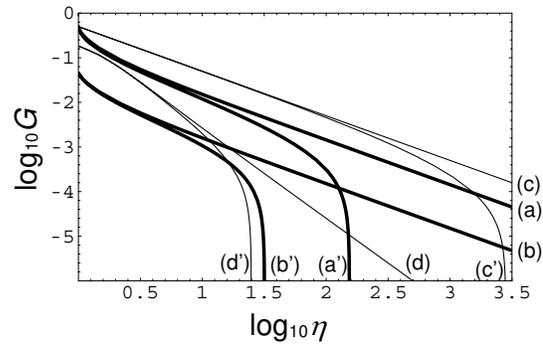}
 \end{center}
 \caption{Key gain $G$ versus transmission $\eta$. 
(a) $\zeta=\gamma=0$. (b) $\zeta=3\%$, $\gamma=0$.
(c) BB84 with an ideal single-photon source
for no errors. (d) BB84 with a coherent-state 
source for no errors. (a')--(d'): The same except for 
inclusion of dark counting rate $\gamma=10^{-4}$.
\label{keyrate}}
\end{figure}

In order to achieve a high key gain, we can optimize
over $|\alpha|^2$ for a given model of errors.
Here we assume that all errors are spurious countings
(curve A), and take 
$\lambda=\gamma+1-\exp\{-|\alpha|^2\eta[4\zeta/(1-2\zeta)]\}$.
The first term is the contribution independent of $|\alpha|^2$,
such as the dark counting rate of the detector. The rest represents
 ``misalignment errors'', which are caused by a stray light proportional 
to the strength of LO. A mode mismatch between Alice's and Bob's LO
is an example of this type of errors. We chose the parameter $\zeta$
such that $n_{\rm err}/n_{\rm fil}\rightarrow \zeta$ for 
$\eta\rightarrow 0$ when $\gamma=0$. Assuming this model, we optimized
$G$ over $|\alpha|^2$, which is shown in Fig.~\ref{keyrate}.
For $\gamma=\zeta=0$ [curve (a)], the key gain decreases as $G\sim O(\eta)$,
which should be compared to the $O(\eta^2)$ decrease in the  
case \cite{ILM01} where a coherent-state source 
is simply substituted for a single-photon source in BB84 [curve (d)].
When $\eta$ is small, the optimal choice is $|\alpha|^2\sim 0.23$,
which gives $n_{\rm fil}/N\sim 0.91(\eta/2)$. The raw key is   
shorten by factor $\sim 0.69$, leading to $G\sim 0.29(\eta/2)$.
This value is smaller than the ideal BB84 $G=\eta/2$ [curve (c)] 
by a constant factor. If we include a small alignment error
$(\zeta=3\%)$, the key rate drops by a constant factor but
 the $O(\eta)$ dependence remains [curve (b)]. This tendency 
continues up to $\zeta\sim 7.6\%$, at which the key gain is zero 
for any value of $|\alpha|^2$. Finally, if we include 
a contribution of dark counting $\gamma$, each curve drops to zero
when the overall counting rate is comparable to $\gamma$.

In summary, we have shown that by encoding on the 
phase of a weak coherent pulse relative 
to a strong reference pulse, we can 
achieve a key rate of $O(\eta)$ with unconditional 
security, which is an advantage over the coherent-state 
BB84. There are several proposals \cite{BB84mod} to improve 
the performance of the coherent-state BB84, 
and their unconditional security is an 
interesting problem. The security of the original B92,
which uses only one LO, is also interesting since the 
relation between the amplitude of the reference pulse 
and the security will show up more tightly.

The author thanks N.~Imoto, H.-K.~Lo, 
D.~Mayers, J.~Preskill, K.~Tamaki, 
 and especially N.~L\"utkenhaus for helpful discussions.

\end{document}